\documentclass[final,times,12pt]{elsarticle}
\usepackage[a4paper, total={6.5in, 9in}]{geometry}

\usepackage{amssymb}
\usepackage{amsmath}
\usepackage[hidelinks]{hyperref}
\hypersetup{
    colorlinks=true,
    linkcolor=black,
    filecolor=magenta,      
    urlcolor=blue,
}
\usepackage{breakurl}
\usepackage{xurl}
\usepackage{xcolor}
\usepackage{tabularx}
\usepackage{tcolorbox}
\usepackage{multirow}
\usepackage{booktabs}
\definecolor{backcolour}{rgb}{0.95,0.96,0.96}
\usepackage{listings}
\lstdefinestyle{mystyle}{
    backgroundcolor=\color{backcolour},   
    numberstyle=\tiny\color{codegray},
    captionpos=b,                    
    keepspaces=true,                 
    columns=fullflexible,
    frame=single,
    breaklines=true,
    showtabs=false,                  
    tabsize=2
}
\lstdefinelanguage{js}{
  keywords={typeof, new, true, false, catch, function, return, null, catch, switch, var, const, if, in, while, do, else, case, break},
  keywordstyle=\color{blue}\bfseries,
  ndkeywords={div, script, /div, /script, class, export, boolean, throw, implements, import, this},
  ndkeywordstyle=\color{darkgray}\bfseries,
  identifierstyle=\color{black},
  sensitive=false,
  comment=[l]{//},
  morecomment=[s]{/*}{*/},
  commentstyle=\color{purple}\ttfamily,
  stringstyle=\ttfamily,
  morestring=[b]',
  morestring=[b]"
}
\usepackage{lineno}

\journal{Journal of Pathology Informatics}

\usepackage[acronym,automake]{glossaries}

\makeglossaries
\renewcommand{\glossaryentrynumbers}[1]{}
\renewcommand{\glossarysection}[2][]{}
\newacronym{wsi}{WSI}{whole slide image}
\newacronym{wsv}{WSV}{whole slide viewer}
\newacronym{idc}{IDC}{imaging data commons}
\newacronym{pacs}{PACS}{picture archiving and communication system}
\newacronym{som}{SoM}{system on module}
\newacronym{ife}{IFE}{Iris File Extension}
\newacronym{html}{HTML}{HyperText Markup Language}
\newacronym{http}{HTTP}{Hypertext Transfer Protocol}
\newacronym{js}{JS}{JavaScript}
\newacronym{jit}{JIT}{just-in-time}
\newacronym{osd}{OSD}{OpenSeaDragon}
\newacronym{api}{API}{application programming interface}
\newacronym{cors}{CORS}{cross-origin resource sharing}
\newacronym{fifo}{FIFO}{first-in-first-out}
\newacronym{tcp}{TCP}{transmission control protocol}
\newacronym{tls}{TLS}{transport layer security}
\newacronym{ssl}{SSL}{secure socket layer}
\newacronym{dicom}{DICOM}{Digital Imaging and Communications in Medicine}
\newacronym{wadors}{WADO-RS}{Web Access to DICOM Objects RESTful services}
\newacronym{hipaa}{HIPAA}{Health Insurance Portability and Accountability Act}
\newacronym{tiff}{TIFF}{tagged-image file format}
\newacronym{ims}{IMS}{image management system}
\newacronym{iod}{IOD}{image object definition}
\newacronym{nema}{NEMA}{National Electrical Manufacturers Association}
\newacronym{nas}{NAS}{network-attached storage}
\newacronym{ieee}{IEEE}{Institute of Electrical and Electronics Engineers}
\newacronym{dzi}{DZI}{Deep Zoom image}
\newacronym{json}{JSON}{JavaScript Object Notation}
\newacronym{jpeg}{JPEG}{Joint Photographic Experts Group}
\newacronym{avif}{AVIF}{Alliance for Open Media Video 1 file format}
\newacronym{xml}{XML}{Extensible Markup Language}
\newacronym{iec}{IEC}{International Electrotechnical Commission}
\newacronym{i2s}{I2S}{Iris Interoperability Standard}
\newacronym{icc}{ICC}{International Color Consortium}
\newacronym{ascii}{ASCII}{American Standard Code for Information Interchange}
\newacronym{utf8}{UTF-8}{Unicode Transformation Format, 8-bit}
\newacronym{sof}{SOF}{start of file}
\newacronym{eof}{EOF}{end of file}
\newacronym{tpt}{TPT}{time per tile}
\newacronym{fov}{FOV}{field of view}
\newacronym{nci}{NCI}{National Cancer Institute}
\newacronym{uid}{UID}{unique identifier}
\newacronym{cdn}{CDN}{content delivery network}
\newacronym{slsa}{SLSA}{Supply-chain Levels for Software Artifacts}

\begin{document}

\begin{frontmatter}


\title{Iris RESTful Server and IrisTileSource: An Iris implementation for existing OpenSeaDragon viewers}

\author[MM]{Ryan Erik Landvater MD MEng \fnref{corAuthor}}
\author[MM]{Navin Kathawa}
\author[MM]{Mustafa Yousif MD}
\author[MM]{Ulysses Balis MD}
\affiliation[MM]{organization={University of Michigan Medical School, Department of Pathology},
            addressline={2800 Plymouth Road},
            city={Ann Arbor},
            postcode={48109-2800},
            state={MI},
            country={USA}
}
\fntext[corAuthor]{Corresponding Author}

\begin{abstract}
The Iris File Extension (IFE) is a low overhead performance-oriented whole slide image (WSI) file format designed to improve the image rendering experience for pathologists and simplify image management for system administrators. However, static hypertext transfer protocol (HTTP) file servers cannot natively stream subregions of high-resolution image files, such as the IFE. The majority of contemporary WSI viewer systems are designed as browser-based web applications and leverage OpenSeaDragon as the tile-based rendering framework. These systems convert WSI files to Deep Zoom Images (DZI) for compatibility with simple static HTTP file servers. To address this limitation, we have developed the Iris RESTful Server, a low-overhead HTTP server with a RESTful API that is natively compatible with the DICOMweb WADO-RS API. Written in C++ with Boost Beast HTTP and Asio networking libraries atop the public IFE libraries, the server offers both security and high performance. Testing shows that a single Raspberry Pi equivalent system can handle a peak of 5,061 req/s (average 3,883 req/s) with a median latency of 21 ms on a private (i.e. hospital) network. We also developed and merged a new OpenSeaDragon TileSource, compatible with the Iris RESTful API, into the next OpenSeaDragon release, enabling simple and immediate drop-in replacement of DZI images within WSI viewer stacks. Designed as a secure cross-origin resource sharing microservice, this architecture includes detailed deployment instructions for new or existing WSI workflows, and the public examples.restful.irisdigitalpathology.org subdomain is provided as a development tool to accelerate WSI web viewer development. All relevant Iris software is available under the open-source MIT software license.
\end{abstract}


\end{frontmatter}



\section{Introduction}
\begin{figure}[t]
    \centering
    \includegraphics[width=1\linewidth]{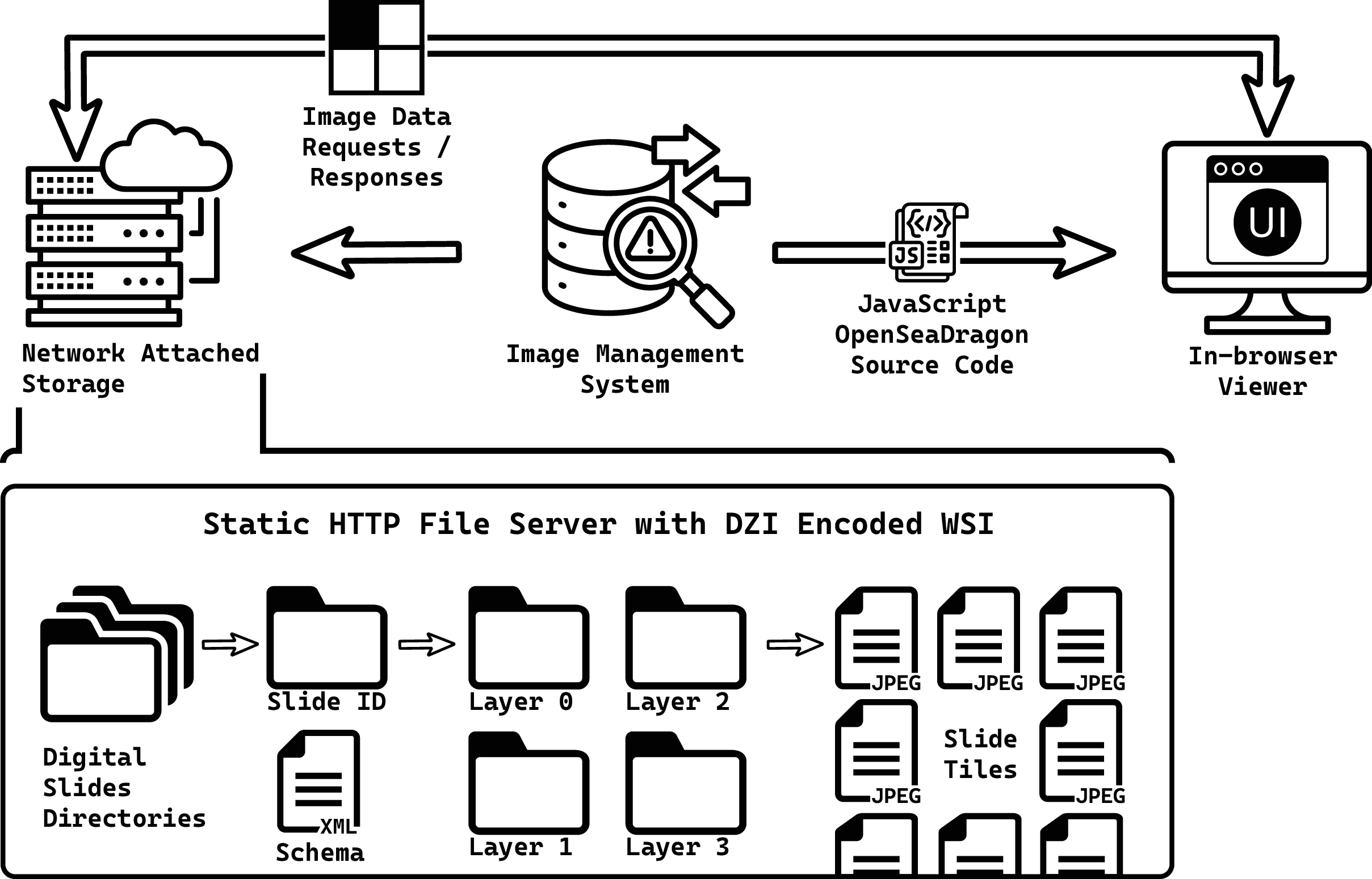}
    \caption{Current \gls{wsi} digital pathology workflow. Digital slides are stored within a \gls{nas} system curated by an \gls{ims} such as a \gls{pacs}. The \gls{ims} often provides both file curation with image retrieval and an \gls{wsv} application by issuing \gls{osd}-based viewer source code to a browser on the pathologist client machine. The \gls{osd} \gls{wsv} application requests regions of digital slide based on the current view port magnification and location on the slide. \Gls{osd} viewer applications historically use \gls{dzi} for native static \gls{http} file server compatibility and are stored as nested directories of individual image files, each encoding a single tile.}
    \label{fig:cur-schema}
\end{figure}

Digital pathology workflows, specifically \gls{wsi} viewer systems, are being rapidly incorporated into a growing number of clinical practices in both academic and community pathology. These systems, broadly, involve a remote \gls{wsi} file repository in the form of \gls{nas} and a \gls{wsv} application that retrieves a portion of the image data from the \gls{nas} via a file server instance. The \gls{wsv} then renders the image portion for pathologist viewing. Viewer systems are generally considered to be one of two types: 1) locally installed applications on the client (pathologist) workstation\cite{Landvater2025}\cite{Bankhead2017} or 2) browser-based web applications\cite{Schffler2022}\cite{Gorman2023}, which are loaded in the form of \gls{html} and \gls{js} source code from a server and run within the browser sandbox. Many \gls{ims} solutions, such as hospitals' \glspl{pacs}, are responsible for curating the collection of digital slides within the \gls{nas} and have historically involved monolithic server deployments\cite{Kawa2022} that include browser-based \gls{wsv} applications (the second variety of the two described above). The general schema for this type of digital pathology workflow is shown in Figure \ref{fig:cur-schema}.

Advancements in the underlying rendering engines promise improved viewer responsiveness for both installed\cite{Landvater2025} and browser-based \gls{wsv} applications\cite{Schffler2022}. The latter variety are aided by \gls{jit} \gls{js} compilers for responsiveness closer to native code\cite{Serrano2021}, but require simultaneous advancements in the methods we employ to access remote slide image data for these improvements to be realized. Current \gls{js} based viewers leverage \gls{osd} as a super-resolution tiled rendering framework\cite{openseadragon}. \Gls{osd} based viewer systems historically use the \gls{dzi} file format, which stores \gls{wsi} files as a markup file -- such as \gls{xml} or \gls{json} -- and a corresponding set of nested directories, each containing a single resolution layer within a multi-resolution image pyramid. Within each of these subdirectories are individual image files corresponding to a single rendered tile (Figure \ref{fig:cur-schema}, bottom pane). For more information on the \gls{osd} pyramidal rendering techniques for super-resolution images, we suggest the \gls{osd} rendering description by Sh\"uffler \textit{et al}\cite{Schffler2021}.

The \gls{dzi} extension cleverly formats a high-resolution image in a structure natively compatible with any simple static \gls{http} file server. Unfortunately, this undesirably incorporates the operating system's file system as a critical internal component of the slide rendering software implementation. Additionally, this generates an expansive set of directories and subdirectories with which a file server's file system and the system administrators must contend. Finally, conversion to \gls{dzi} from the original format, such as the \gls{dicom} format or a proprietary slide scanner format, can be onerous, time consuming, and result in loss of clinical metadata. To address some of these issues, an \Gls{osd} TileSource for digital pathology was developed to improve \gls{wsv} performance by using the OpenSlide\cite{Goode2013} \textit{DeepZoomGenerator} extension to convert to non-standard \gls{dzi} formatted tiles with flexible tile sizes and layer magnifications instead of the sequential 2x-downsampling used in \gls{dzi} conversion \cite{Schffler2021}. However, it would be preferable to avoid the overhead of \gls{dzi} conversion by issuing \gls{wsi} tile data natively from a \gls{wsi} file format.

We recently introduced the \gls{ife} as a vendor-neutral open-source binary container file specification explicitly designed for performance-oriented \gls{wsi} viewer systems that supports modern compression, a dynamic file structure, deep file validation routines, corruption recovery, and annotations\cite{Landvater2025-2}. However, static \gls{http} file servers can only transmit complete files -- or for large files like \glspl{wsi}, must stream elements using specialized client \gls{js} code able to decode partial reads (`Range Requests' with \gls{http} 206 response). \gls{ife}, via the Iris Codec library, supports this type of advanced client-side WebAssembly module\cite{Landvater2025-2} when files are hosted on static \gls{http} servers that can process range requests (like Amazon S3); however, such an implementation requires complex \gls{js} code and is still not as efficient as a server with native \gls{wsi} streaming capabilities tailored to the file structure. A simple \gls{http} RESTful request structure is therefore far preferable. Herein we introduce the Iris RESTful server, an open-source \gls{ife} compatible \gls{http} server implementation with both its own RESTful \gls{api} and compatibility with the \gls{dicom}web's \gls{wadors} \gls{api} for ease of interoperability. 

The Iris RESTful Server is written in C++ using the low-level and highly templated Boost Beast \gls{http}\cite{boostbeast} and Asio networking libraries, and is built on the open-source \gls{ife} libraries that we have made publicly available\cite{irisfileextensionrepo}. The Iris RESTful Server was designed to be deployed as a modular and scalable containerized microservice accessible via a separate \gls{cors} domain to integrate with current \gls{ims} \gls{wsv} applications. We also developed and merged a new \gls{osd} TileSource, the \textit{IrisTileSource}, into the next \gls{osd} release so that \gls{ife} encoded files can be immediately and seamlessly incorporated into existing digital pathology workflows. Incorporating \textit{IrisTileSource} requires changing four lines of code within a viewer source (see Code listing \ref{lst:html} below). Finally we make openly available the \url{https://examples.restful.irisdigitalpathology.org/} subdomain, which we refer to as the \textbf{Iris RESTful Example domain}, as a fully Iris RESTful \gls{api} compliant server with example \gls{ife} encoded slides for community use as a tool to aid in streamlining the integration of \gls{ife} encoded slides.

\section{Design}
\subsection{Iris RESTful Server}
\begin{figure}[t]
    \centering
    \includegraphics[width=1\linewidth]{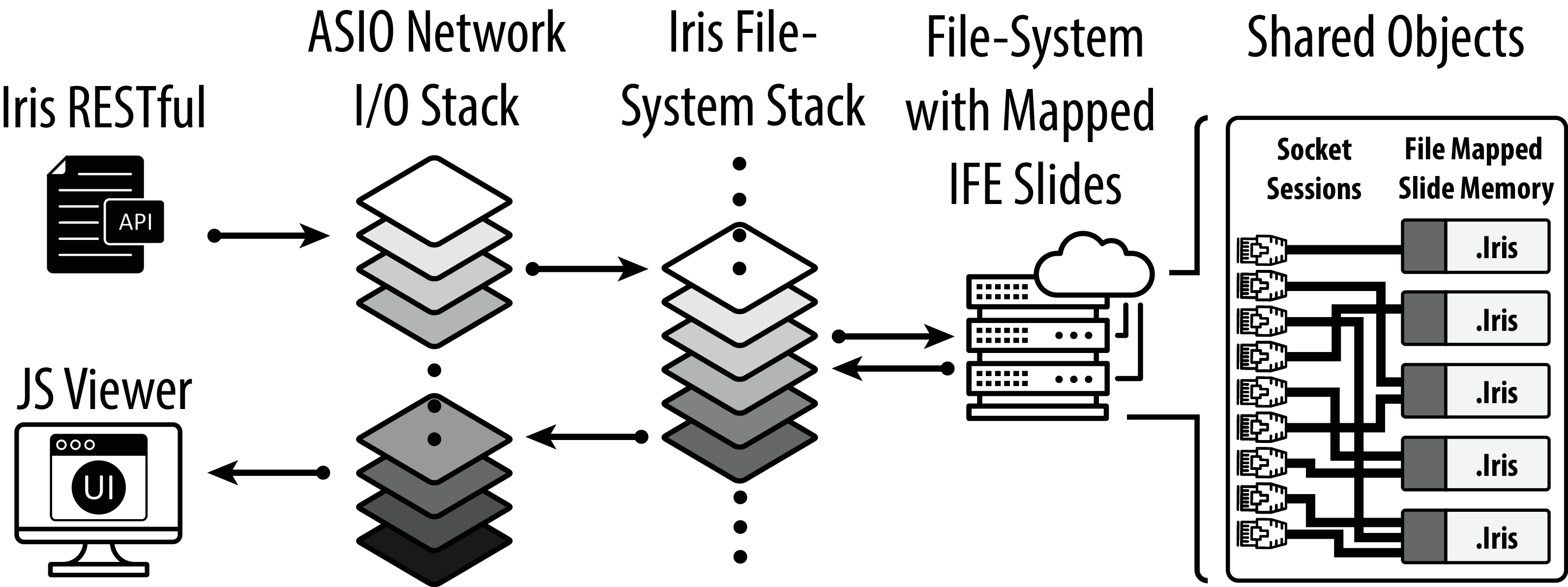}
    \caption{Iris RESTful server architecture. Isolated stacks execute the server requests on multiple independent threads in \gls{fifo} ordering. Asio Networking reactor threads are solely responsible for accepting packet requests and writing responses. The task of interpreting requests is forwarded to a separate Iris lockless file system stack and executed on a configurable number of separate independent threads. This reduces the pressure on the networking stack, limiting the network threads' scope to the execution of networking kernel calls in order to speed the draining of the network buffer and ensure high-reliability. File system kernel calls are isolated to the Iris stack, where slides are mapped into virtual memory space and shared between \gls{tcp} socket sessions accessing the same slide resources. }
    \label{fig:server}
\end{figure}

The Iris RESTful server was designed as an extremely high-performance, lightweight, and secure dedicated slide-serving microservice. It is configured to scale independently of the \gls{ims} and only returns slide tiles in response to Iris RESTful \gls{api} requests -- though the server may be optionally configured to perform static file services such as providing an \gls{osd} viewer web-application. The server comprises two independently operating stacks: 1) the networking stack and 2) the file system stack (Figure \ref{fig:server}). The server has security logic that parses \gls{api} requests and discards requests that do not adhere strictly to the \gls{wadors} or Iris RESTful \gls{api} (Code listings \ref{lst:API}). When the server is configured as a static \gls{http} web server, it may also stream \gls{dzi} image tiles for dual \gls{ife} and \gls{dzi} compatibility; however, it will only return known file types and limit the file directory scope to ensure continued system security. Furthermore, the system automatically configures \gls{tls} connections for end-to-end encryption when deployed with application load balancers to fulfill the in-transit \gls{hipaa} requirement.

\begin{lstlisting}[basicstyle=\small,caption={Iris RESTful API, acceptable GET commands. These URL target sequences retrieve slide metadata (in JSON form) and compressed slide tile bytestreams, respectively, using the Iris RESTful (\textit{top}) and WADO-RS (\textit{bottom}) APIs}, label={lst:API}, captionpos=b, float]
 Iris RESTful
 GET <URL>/slides/<slide-name>/metadata
 GET <URL>/slides/<slide-name>/layers/<layer>/tiles/<tile>

 Supported WADO-RS (DICOMweb)
 GET <URL>/studies/<study>/series/<UID>/metadata
 GET <URL>/studies/<study>/series/<UID>/instances/<layer>/metadata
 GET <URL>/studies/<study>/series/<UID>/instances/<layer>/frames/<tile>
\end{lstlisting}

As mentioned above, the system has two core stacks (Figure \ref{fig:server}). Each stack is composed of a \gls{fifo} queue with multiple independent computer threads that add to and remove from the queue in a concurrent lockless manner. The division of the server implementation into two separate stacks uncouples networking calls from file system calls. This avoids prolonged computations or tasks from slowing down the server responsiveness to network requests; thus, one or more prolonged requests are isolated to a limited number of threads and passed over by the rest of the server. The Asio networking stack is solely dedicated to clearing the network buffer as quickly as possible to avoid dropping packets, thus increasing the server reliability and throughput. 

The server is built on top of the public \gls{ife} (de)serialization library\cite{Landvater2025-2}. When a client issues a new \gls{http} request, an Asio networking thread performs \gls{tcp} connection procedures, executes a \gls{tls}/\gls{ssl} handshake, and establishes a new secure \gls{tcp} socket session. After validating the request format, the thread passes the request to the file system \gls{fifo} queue for interpretation. A file system thread looks into the server's active \gls{ife} slide handles and if the slide is not presently open, validates the slide metadata against the \gls{ife} specification. It then creates a new \gls{ife} slide handle if the requested \gls{ife} file is present and has passed the \gls{ife} validation. In this way, multiple client sockets may benefit from shared \gls{ife} memory-mapped resources (Figure \ref{fig:server}, right-side). \gls{ife} slide handles are reference-counted server objects; they persist so long as at least one \gls{tcp} socket remains open with a client viewer. When the last client socket referencing an \gls{ife} handle closes, the slide is unmapped from virtual memory space. During a request and after slide information is abstracted from the \gls{ife} slide handle, the slide data is serialized into a \gls{http} response and passed back to the Asio networking stack for transmission to the \gls{wsi} viewer. 

\subsection{OSD IrisTileSource}
The \textit{IrisTileSource}, a subclass derived from the \gls{osd} TileSource class, is designed to work seamlessly with the Iris RESTful Server \gls{api}. The metadata and retrieve-tile overloaded functions are the main mechanisms by which the \textit{IrisTileSource} class interfaces with a server instance via the Iris RESTful \gls{api} (Code listings \ref{lst:API}). The slide metadata is retrieved and parsed, and the \textit{IrisTileSource} configures the \gls{osd} viewer instance by iterating over each layer tiles-extent and scale parameters. Viewer configuration is automatic. Subsequent tile requests follow a similar structure to \gls{dzi} calls, wherein the tile location in $x$ and $y$ coordinates of layer $l$ are converted into a URL-target sequence with a layer tile index $t$ following a raster pattern according to the following equation (\ref{eq:tile-index}):
\begin{equation}
\label{eq:tile-index}
    t = y * tiles_{l,x} + x
\end{equation}
where $tiles_{l,x}$ is the total number of tiles in the $x$ dimension of layer $l$. A tile URL-target is thus generated per tile requests with the following structure:
\begin{lstlisting}
    GET <server-domain>/slides/<slide-name>/layers/l/tiles/t
\end{lstlisting}

\section{Implementation}
\subsection{Iris RESTful Server}
\begin{figure}[t]
    \centering
    \includegraphics[width=1\linewidth]{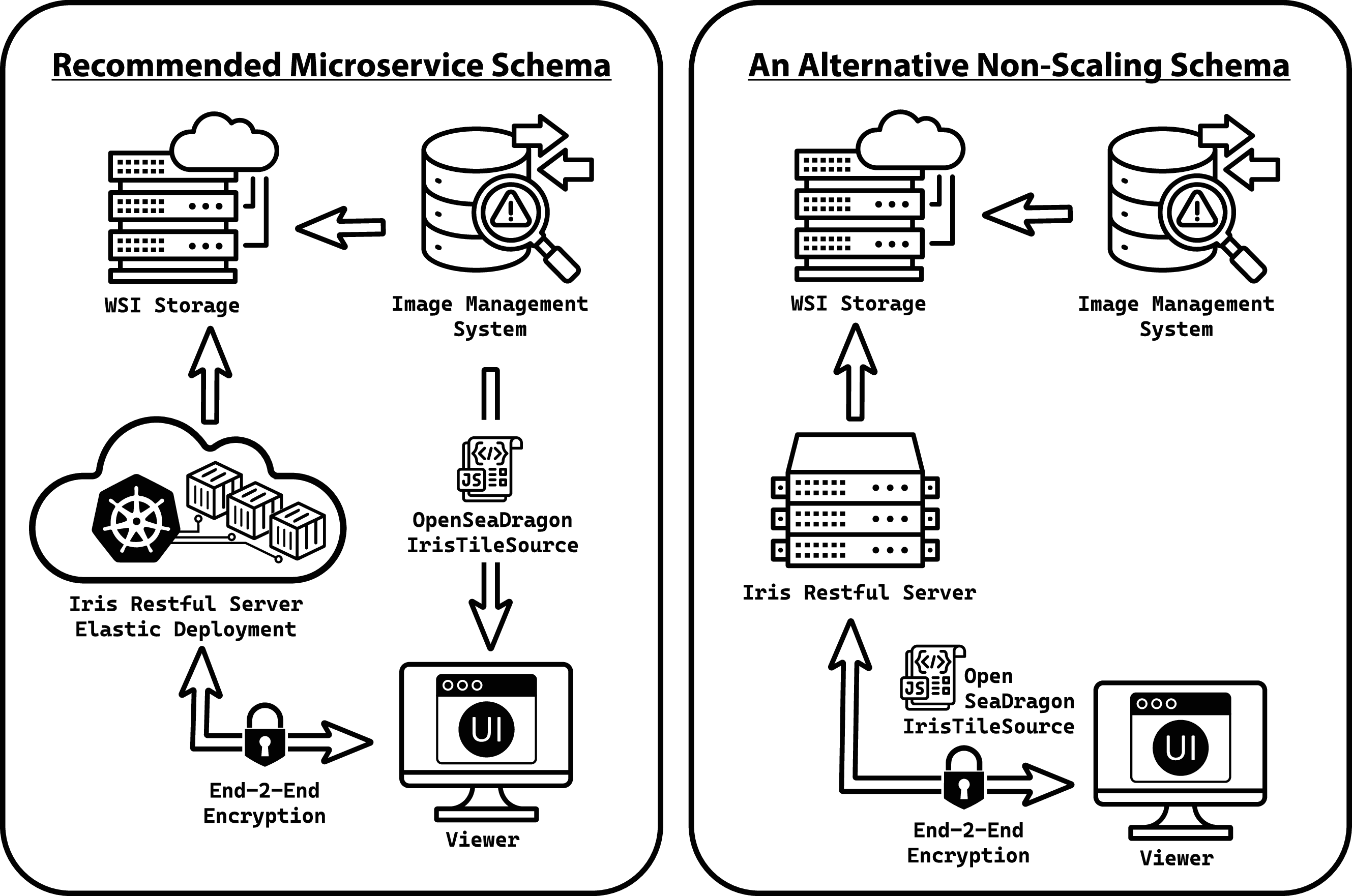}
    \caption{Iris RESTful within digital pathology workflows. (\textit{left}) Our suggested architecture separates the \gls{ims} and \gls{wsv} application from a scalable RESTful instance on a container orchestrator like Kubernetes (or AWS Elastic Container Service / Google Run). In this design, Iris RESTful is a \gls{cors} domain solely responsible for issuing slide tiles from the \gls{nas}. (\textit{right}) In a different implementation, Iris RESTful is hosted on a server without container orchestration (but may be in or outside of a Docker container) with static \gls{http} server capabilities enabled -- allowing it to issue the \gls{osd} based \gls{wsv} application directly. These are two opposite example architectures; however, there are many other viable workflow designs.}
    \label{fig:schema}
\end{figure}

The server can be implemented as either a simple runtime executable or within a containerized environment. The server responds to both the Iris RESTful API and \gls{wadors} API, shown in Code listing \ref{lst:API}. The full source code for the RESTful server with build configuration files are available on our Github Repository and can be compiled for any architecture; however, we recommend deployment as a elastic / scalable container service using the official Docker container images we host on our Github Container Repository. Please refer to the Data Availability section for more information on acquiring the source code and binary distributions. The proposed digital pathology workflow architectures are illustrated in Figure \ref{fig:schema}, with our suggested scalable container architecture on the left. A simplified single instance server with mixed Iris RESTful and static \gls{http} file serving capabilities -- which allows an Iris RESTful server to independently host a \gls{wsv} application -- is illustrated on the right. We will describe the steps for implementing the server below, starting with a basic example and expanding to the more advanced configuration. 

In the most basic implementation, a server instance can be launched in a single line of bash code using a Docker run command (Code listing \ref{lst:run}). Alternatively, the service can be run outside of a containerized instance by downloading one of the pre-compiled release binaries from the Github \href{https://github.com/IrisDigitalPathology/Iris-RESTful-Server/releases/latest}{Iris-Restful-Server Releases} page and configuring the server using the arguments in Code listing \ref{lst:run}. When run in a container, the server is automatically configured with default arguments; however, these may be overridden using the Docker conventional methods for defining \lstinline{CMD} parameters (see Docker documentation). A slide directory containing the \gls{ife} encoded digital slide files is a required argument (Code listing \ref{lst:run}, \textit{argument -d / --dir}). To activate the static \gls{http} file serving capabilities, a document root directory may be specified (Code listing \ref{lst:run}, \textit{argument -r / --root}). It should be noted that Iris will prohibit clients from downloading \gls{ife} files even in this mode; therefore, you may safely contain slide files in the same directory as \gls{dzi} or other permitted files for static file service without risking client \gls{ife} file download access. If Iris is configured for static file serving and is providing the \gls{wsv} application, the \gls{cors} (cross-domain security policy specified with the '-o/--origin' argument) can be disabled since the Iris RESTful is both the domain for the \gls{wsv} application and the source of the slide tiles. This description corresponds with the right pane of Figure \ref{fig:schema}.

\begin{lstlisting}[basicstyle=\small, caption={An example Docker deployment statement to begin running an Iris RESTful instance. A directory containing the IFE encoded slide files must be mounted to the container. In Docker convention, arguments for the containerized application may follow the Docker image name and these optional arguments for Iris RESTful are shown below. For a more detailed explaination visit the Iris-RESTful-Server Github Repository.}, label={lst:run}, captionpos=b, float, floatplacement=t]
docker run --rm -v${slide-dir}:/slides -v${doc-root}:/doc-root -p${port}:3000 ghcr.io/irisdigitalpathology/iris-restful:latest -d /slides -p 3000 -r /doc-root <addn-args>

Where the following arguments are understood by Iris RESTful:
-h --help: Print this help text 
-p --port: System port on which to listen for client connections.
-d --dir: Path to the directory containing the Iris Slide Files to be served
-c --cert: Public SSL certificate in PEM format for establishing HTTPS connections
-k --key: Private key in PEM format to sign argument provided in CERT
-o --cors: Slide viewer domain. Returned in `Access-Control-Allow-Origin' header
-r --root: Static server document root directory for activating RESTful server as file server
--http-only --no-https: Disable TLS / SSL layer. Server will respond to HTTP not HTTPS.
If run without defining the -r/--root option, HTTPS responses will contain 'Access-Control-Allow-Origin':`*' wild-card unless the `-o/--cors option' is defined as the IMS domain. 
\end{lstlisting}

In a more scalable configuration, a cloud cluster service can be deployed using a container orchestrator system such as Kubernetes or Kubernetes-derived services such as those by Amazon Web Service (Elastic Container Service) or Google Cloud (Cloud Run). This design delegates container scaling to a separate hypervisor system, and is more appropriate for large-scale implementations, such as institutions with dedicated hospital information technology offices familiar with scalable container orchestration services. Although static file serving capabilities can be enabled in this configuration (this is the architecture of examples.restful.irisdigitalpathology.org), for this type of production workflow we recommend using the server as a \gls{cors} domain where the \gls{ims}/\gls{pacs} service (such as SECTRA AB or Roche) provides the \gls{wsv} application from their \gls{pacs} server instance and assigns the Iris-RESTful implementation as the \gls{osd} `serverUrl' (see OSD \textit{IrisTileSource} implementation Section \ref{sec:implementOSD-ITS} below). When deployed in this configuration, the RESTful server automatically generates self-signed 2048-bit RSA certificates that it uses to coordinate a secure connection with the container orchestrator / load balancer to ensure end-to-end encryption, even within a hospital's secure network or a provider virtual private network, fulfilling the in-transient \gls{hipaa} encryption requirement. 

Please refer to the online documentation within the Iris-RESTful-Server Github repository for a more thorough and up-to-date explanation of server deployment strategies. 

\subsection{OSD IrisTileSource}
\label{sec:implementOSD-ITS}
The \textit{IrisTileSource} can be implemented within existing viewers by substituting only four lines of \gls{js} code in the \gls{osd} viewer class declaration. Rather than providing an \gls{osd} viewer ``tileSources'' parameter with the path to \gls{dzi} image file, you must configure the \gls{osd} viewer ``tileSources'' as the new derived \textit{IrisTileSource} class and assign that class the ``serverUrl'' parameter with the Iris RESTful server domain as well as the \gls{ife} encoded slide file name (excluding the `.iris' extension). See example \gls{html} in Code listings \ref{lst:html} for a simple functional example web-viewer application.

\begin{lstlisting}[language=js, showstringspaces=false, label={lst:html}, caption={Simple HTML source that runs an OSD viewer using the \textit{IrisTileSource} in only four lines of additional code using the latest version of OSD. The new \textit{IrisTileSource} is assigned to the OSD viewer tileSources parameter and requires the new CORS serverUrl and slideID parameters. This example uses the Iris RESTful Example domain as the server and the cervix\_2x\_jpeg.iris example slide file. CORS is enabled using the Anonymous value.}, float]
<div id="viewer" style="width: 100%; height: 100%;"></div>
// For OSD release > v5.0.1
<script src="https://cdn.jsdelivr.net/npm/openseadragon@latest"></script>
// Otherwise you must download a copy and assign that source 
// src="openseadragon/openseadragon.min.js"
    const viewer = OpenSeadragon({
        id:"viewer",
        tileSources: {
            // New IrisTileSource parameters
            type: "iris", // Inform OSD this is an Iris source
            serverUrl: "https://examples.restful.irisdigitalpathology.org",
            slideId: "cervix_2x_jpeg", // For the file "cervix_2x_jpeg.iris"
            crossOriginPolicy: "Anonymous" // Enable CORS for RESTful Example
            // End of new parameters
        }
    });
</script>
\end{lstlisting}

\subsection{Error Handling}
The server is runtime hardened with numerous try-catch blocks that prevent runtime errors from escaping each stack. In the event of invalid requests or server-side errors, the server will terminate parsing only a single request and immediately respond with a series of HTTP error codes and response body narratives explaining the server error if such a typical error exists within the dictionary of known issues (such as failed validation or request range violations). The stack thread will then simply move on in the queue. The major response error categories are available within the \textit{IrisRestfulTypes.hpp} class definitions. All error response descriptions are returned in the HTTP response body in plain-text form for ease of parsing. 

\section{Performance Results}
\subsection{Materials and Methods}
The server distributions were built using GitHub Actions runners directly from the source repository using the latest versions of supported operating systems (GitHub Actions YAML compilation code is available at the repository website). Secondary builds were also performed using Google Cloud Build runners and evaluated for security insights. Vulnerability scans of these automated build artifacts from the source repository (\gls{slsa}\cite{slsa}, Build Level 3) could not identify any vulnerabilities in the resulting server deployment (vulnerabilities ranging from low to critical). 

Illustrative performance testing was conducted for both a single-instance local network deployment (Figure \ref{fig:schema}, \textit{right}) and the public facing Iris RESTful Example domain in comparison to other well-established and publicly available \gls{wsi} servers, specifically Path-Presenter and the \gls{nci} \gls{idc} Slim viewer\cite{Serrano2021}. The slides were selected for similar imaging characteristics (JPEG encoded 256 x 256 pixel tiles) which resulted in similar packet sizes. Iris Restful Example did not use a \gls{cdn}; although we are unable to verify that \gls{cdn} caching was not used by Slim or PathPresenter. Performance was evaluated using Locust\cite{locust}, a widely accepted open-source and flexible load testing tool. Locust explicitly does not allow local data caching during a test. A ramped load test was chosen to allow for automatic instance scaling in the publicly facing \gls{wsi} servers. Users were scaled up to 160 and 220 concurrent users for public \gls{wsi} server and local network tests, respectively. Each virtual user issued a tile request at 1-2 ms intervals after receiving the prior response. Custom locust python scripts were created for each target server and tiles were randomly sampled from within image bounds. These scripts have been made available, along with the raw results of these tests (see Data Availability section). All tests were performed on a 2020 13-in M1 MacBook Pro (Apple, Palo Alto, CA) with 8 GB of RAM running macOS 15.4, with Locust scripts executed using Python 3.11 in an Anaconda environment over a $\geq$ 200 mbps wireless (Wifi) connection. 

The single-instance local network deployment was hosted on Turing RTK1 (Turing Machines Inc, Cupertino, CA) 8-core Rockchip RK3588 \gls{som} circuit board with 8 GB of RAM running Ubuntu 24.04 within a Docker container and with a 1 gigabit Ethernet switch. The Turing RTK1 \gls{som} is similar to the Raspberry Pi CM4 or the Jetson Orin Nx. The publicly facing Iris RESTful Example domain is hosted on AWS using an application load balancer in front of a Fargate elastic container service with 1 virtual CPU and 2 GB of RAM and set to scale-out at either 70\% CPU consumption or greater than 500 requests/sec per instance (up to 100 USD/month with aggressive load testing). Both the local network deployment and Iris RESTful Example domain served \gls{ife} formatted tiles using the Iris RESTful \gls{api}. Path-Presenter served \gls{dzi} formatted tiles using a static \gls{http} server without an \gls{api}. The \gls{nci} \gls{idc} repository served \gls{dicom} formatted frames using the \gls{wadors} \gls{api}.

\begin{table}
    \centering
    \begin{tabularx}{\textwidth}{l X X X X X}
        \toprule
        \multirow{2}{*}{\gls{wsi} Server} & \multicolumn{2}{l}{Requests/s} & \multicolumn{3}{l}{Response Times}\\ 
        & Avg & Max & 50\% & 75\% & 95\% \\
        \midrule
        Iris Private Instance & 3883 & 5061 & 21 ms & 33 ms & 63 ms\\
        Iris Restful Example & 1241 & 1944 & 59 ms & 71 ms & 120 ms\\
        Path-Presenter (DZI) & 987 & 1560 & 79 ms & 93 ms & 160 ms\\
        Slim (DICOMweb) & 114 & 146 & 650 ms & 1.2 sec & 2.3 sec\\
        \bottomrule
    \end{tabularx}
    \caption{Aggregate ramped Locust server load testing performance metrics. Iris Private Instance shows ramped load testing for a single Iris RESTful deployment on a local private network. Iris RESTful Example, Path-Presenter, and Slim show aggregate ramped load testing for publicly available \gls{wsi} servers issuing \gls{ife}, \gls{dzi}, and \gls{dicom} slides, respectively.}
    \label{tab:performance}
\end{table}

\begin{figure}
    \centering
    \includegraphics[width=1\linewidth]{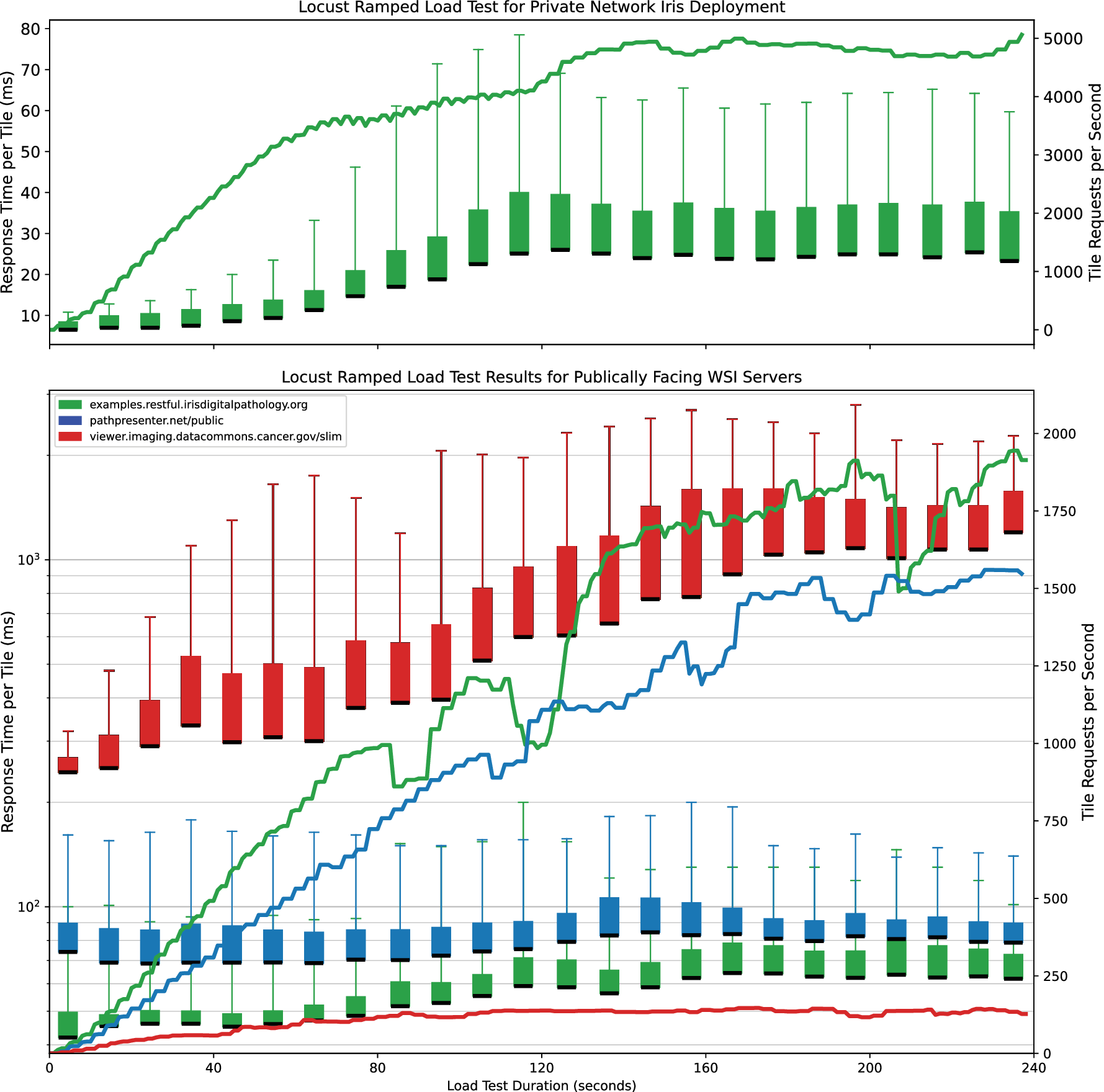}
    \caption{Ramped Locust server load testing performance. An ideal server shows low response time (boxes) while handling a high tile request rate (line trace). (\textit{top}) Ramped load testing for a single Iris RESTful deployment on a local private network without container orchestration with 220 virtual users ramped at a rate of 1.8 usr/sec. The half-box plots show median / 50\%-ile (bottom black line) to 75\%-ile (top) and 95\%-ile (whiskers) server response latency. The line trace shows total tile requests per second. (\textit{bottom}) Comparisons of ramped load testing on publicly available \gls{wsi} servers with 160 virtual users ramped at a rate a 1 user per second showing both tile response times (box-plots) and tile requests per second (line-trace) for Iris RESTful example server (green), Path-Presenter (blue), and \gls{nci} \gls{idc} Slim viewer (red). A log scale was used for tile response times in the bottom pane to aid in visualization due to substantially prolonged \gls{nci} \gls{idc} (\gls{dicom}web) response times.}
    \label{fig:performance}
\end{figure}

\subsection{Server Load Testing}
Results from the local network deployment load tests (n = 932269 requests) showed a median tile response time of 21 ms (range 2.7 ms - 450 ms) with an average tile request rate of 3883 requests/sec over the entire test duration (Figure \ref{fig:performance}, \textit{top pane}). The public facing and scalable Iris RESTful Example domain load tests (n = 298002 requests) showed a median response time of 59 ms (range 34.7 ms - 2.59 sec) with an average tile request rate of 1241 requests/sec over the test duration (Figure \ref{fig:performance}, \textit{bottom pane, green}). Path-Presenter domain load tests (n = 237102 requests) showed a median response time of 79 ms (range 56.8 ms - 1.74 sec) with an average tile request rate of 987 requests/sec over the test duration (Figure \ref{fig:performance}, \textit{bottom pane, blue}). The Slim viewer domain load tests (n = 27392 requests) showed a median response time of 650 ms (range 159.0 ms - 23.7 sec) with an average tile request rate of 114 requests/sec over the test duration (Figure \ref{fig:performance}, \textit{bottom pane, red}). The complete results are shown in Table \ref{tab:performance}.

\section{Discussion}
We previously published on the rapid encoding speeds of \gls{ife} files from source \gls{wsi} files such as \gls{dicom} due to the highly parallelized encoding capabilities afforded by the layer-agnostic tile ordering within the file specification\cite{Landvater2025-2}. This aids in the value of the Iris RESTful server by allowing rapid conversion from source \gls{dicom} to \gls{ife} files, which may then be served to \gls{osd} viewers. The current versions of the Iris Codec \gls{ife} libraries can generate a derived pyramid file (with 2x or 4x downsampling) from a single highest-resolution \gls{dicom} or vendor format source within 1-2 mintues  -- or much faster if OpenSlide decoding\cite{Goode2013} is not needed. We continue to work on improving encoding systems and in future releases will add stream encoding (a slide scanner \gls{api}) to allow for simultaneous generation of archive \gls{dicom} and \gls{ife} files directly from the scanner to make \gls{ife} slides available to Iris RESTful in \gls{wsi} workflows immediately. The \gls{ife} is in the public domain under a Creative Commons Attribution-No Derivative 4.0 license, and Iris Codec and IrisRESTful server implementations are available to the community under the MIT software license. The IrisTileSource is a part of the \gls{osd} repository and follows the \gls{osd}`s BSD-3 license.

In this paper we performed stressful load testing on \gls{wsi} systems to assess system durability and performance when these servers. This includes both expected daily loads early in the load test, as well as abnormally high volumes of requests. The early metrics better align with what we have previously observed with the Slim viewer of an approximately 200 ms median response time (Figure \ref{fig:performance}, \textit{left-side}) in our prior publication\cite{Landvater2025-2}. The Iris RESTful Example domain response times of approximately 20-30 ms also align closely with what we experienced when using our example domain during the development of the \textit{IrisTileSource} and example browser-based \glspl{wsv}. 

We are particularly happy with the server responsiveness on small and low-cost single-board computers such as the Turing RTK1 (which costs only \$200 USD) and on 1 vCPU serverless AWS (Fargate) instances. The \gls{ife} implementations are designed with computationally trivial validations and low instruction set reads, while the Boost Beast library almost exclusively uses inline template functions to create a very lean and light server. The implications of this design include substantially lower hardware cost for \gls{wsi} servers. Implementing a digital pathology workflow can be quite expensive. While the cost of the \gls{wsi} server is just a fraction of the overall cost, it can represent an significant contribution. Leaner and more efficient systems benefit all potential health systems. We are particularly hopeful that with this performance on hardware carrying a smaller financial footprint, we hope that Iris RESTful can help to ensure globally accessible first-class digital pathology workflows for health systems regardless of the country or region's resource status.

\begin{figure}
    \centering
    \begin{tcolorbox}[
    colback=white,
    colframe=black,
    title=Software Accessibility Summary,
    fonttitle=\bfseries\Large,
    boxrule=0.8pt,
    arc=0pt,
    width=\textwidth
  ]
  Code and Binary Artifact Hyperlinks:\\
    \href{https://github.com/IrisDigitalPathology/Iris-RESTful-Server}{RESTful Server Source Code}\\
    \href{https://ghcr.io/irisdigitalpathology/iris-restful:latest}{RESTful Server Containers/Packaged Releases}\\
    \quad\href{https://github.com/IrisDigitalPathology/Iris-RESTful-Server/releases/latest}{RESTful Server Binary Releases}\\
  \href{https://github.com/openseadragon/openseadragon/blob/master/src/iristilesource.js}{OpenSeaDragon IrisTileSource}\\
  \href{https://examples.restful.irisdigitalpathology.org}{Iris RESTful Example domain}
  \end{tcolorbox}
    \label{table:software}
\end{figure}

\section{Conclusion}
The Iris RESTful Server and \gls{osd} \textit{IrisTileSource} allow for immediate and easy replacement of \gls{dzi} images with \gls{ife} encoded slide files within digital pathology \gls{wsi} workflows. Prepackaged container images allow for server deployment in a scalable microservice architecture and require altering only four lines of viewer code to begin rendering with the new IrisTileSource within existing \gls{osd} viewers. Written in C++ with the highly templated Boost Beast library and on top of the \gls{ife} deserialization library this is a low-overhead fast deployment that can issue slide tiles in 21 ms on average under an high request load of over 3883 tile requests per second, even when hosted on a small \$200 USD compute module chipboard (the equivalent of a Raspberry-Pi CM4). We provide the Iris RESTful Example domain, a server that hosts example \gls{ife} encoded slide files and responds natively to Iris RESTful \gls{api} calls, for community use to evaluate HTTPS implementations with the Iris RESTful API and accelerate the development process. All relevant Iris implementations are available under the MIT software license.

\section{Acknowledgments}
We would like to acknowledge Vinnie Falco, the author of Boost Beast. His library and responses to questions on performance optimization were vital to the development and subsequent improvement of the overall server design. We would also like to thank Ian Gilman, the author of OpenSeaDragon for his numerous insights and suggestions for developing and improving upon the IrisTilesSource, as well as integrating it into the OpenSeaDragon library. We would like to acknowledge Dr. Corey Post for a thorough manuscript review and recommendations.

\section{Author Contributions}
RL designed and developed the Iris RESTful server, Iris RESTful API, performed the performance profiling, created the figures, wrote the introduction, performance results, and conclusion sections, and contributed to the design and implementation sections. NK authored the \textit{IrisTileSource}, integrated it with OpenSeaDragon (including authoring additional documentation), and contributed to the design and implementation sections. MY provided guidance and insights on the structure of the Iris RESTful API and support of the WADO-RS API. UB provided guidance throughout the development. All authors reviewed and approved the final manuscript. 

\section{Data Availability Statement}
The official container images for the Iris RESTful server are available on the Github Container Repository (\url{https://ghcr.io/irisdigitalpathology/iris-restful:latest}). The \gls{js} source code for the \textit{IrisTileSource} is available within the \gls{osd} Github Repository and can be evaluated more closely within \gls{osd} pull request \#2759 (\url{https://github.com/openseadragon/openseadragon/pull/2759}). The complete source code for the Iris RESTful server is available at the official Iris Digital Pathology Github organization page (\url{https://github.com/IrisDigitalPathology}) with build instructions as well as pre-compiled binary artifacts with each release within the releases directory (\url{https://github.com/IrisDigitalPathology/Iris-RESTful-Server/releases/latest}). \gls{ife} encoded slide files are available through the Iris RESTful Example domain (\url{https://examples.restful.irisdigitalpathology.org}) and at \url{https://iris.example-slides.org}. To generate \gls{ife} encoded files, please refer to \gls{ife} online documentation (\url{https://github.com/IrisDigitalPathology/Iris-Codec#implementations}) or review the data availability statement (\url{https://www.sciencedirect.com/science/article/pii/S2153353925000471#da0005}) with the original \gls{ife} publication\cite{Landvater2025-2}. The performance Locust test scripts and raw test results data (csv) are included as supplemental data with this publication.


\appendix
\section{Glossary of Terms}
\printglossaries

 \bibliographystyle{elsarticle-num}
 \bibliography{bibliography.bib}



\end{document}